\documentclass[a4paper]{jpconf}
\usepackage{graphicx}
\usepackage{amsfonts,amsmath}
\begin{document}
 
\title{Horizon effects for surface waves in wave channels and circular jumps}
\author{G. Jannes\footnote{Speaker}, R. Piquet, J. Chaline, P. Ma\"{i}ssa, C. Mathis, G. Rousseaux}
\address{Universit\'{e} de Nice Sophia Antipolis, Laboratoire J.-A. Dieudonn\'{e}, UMR
CNRS-UNS 6621, Parc Valrose, 06108 Nice Cedex 02, France}
\ead{gil.jannes@unice.fr}
\begin{abstract}
Surface waves in classical fluids experience a rich array of black/white hole
horizon effects. The dispersion relation depends on the characteristics of the
fluid (in our case, water and silicon oil) as well as on the fluid depth and the
wavelength regime. In some cases, it can be tuned to obtain a relativistic
regime plus high-frequency dispersive effects. We discuss two types of ongoing
analogue white-hole experiments: deep water waves propagating against a
counter-current in a wave channel and shallow waves on a circular hydraulic jump.
\end{abstract}

\section{Introduction}
Surface waves in classical fluids provide a natural and rich class of black/white hole analogues. Two familiar examples are the blocking of sea waves at a river mouth and the approximately circular jump created by opening the tap in a kitchen sink. We reproduce these two types of white hole analogues in controlled laboratory settings in order to study the associated horizon effects and their possible lessons for relativity (and vice versa: lessons from relativity for fluid mechanics). The river-mouth example corresponds to deep water waves propagating against a
counter-current in a wave channel, while the kitchen-sink example corresponds to shallow waves on a circular hydraulic jump.

The general dispersion relation for capillary-gravity surface waves propagating against a counter-current of velocity $U$ is 
\begin{equation}\label{disp-rel}
 (\omega-Uk)^2=\left(gk+\frac{\gamma}{\rho}k^3\right)\tanh(kH),
\end{equation}
with $g$ the gravitational constant, $\gamma$ the surface tension, $\rho$ the density, $H$ the fluid depth, and as usual $\omega$ and $k$ are the frequency and wavenumer, respectively. By developing the two extreme cases $kH\ll 1$ and $kH \gg 1$ one obtains the deep water and the shallow water limit which are applicable to the wave channel and the circular jump, respectively.

\section{Deep water waves in a wave channel}
The deep water case gives $(\omega-Uk)^2\approx gk+\frac{\gamma}{\rho}k^3$. The gravity wave limit is obtained by neglecting the term in $k^3$, and has a white hole horizon when $U=\sqrt{g/k}$. The deep water case with inclusion of capillarity possesses a tremendously rich phenomenology which includes not only white hole horizons, but also additional horizons such as a negative horizon (a blocking line for waves with a negative co-moving frequency) and a blue horizon (a blocking line for waves blue-shifted due to the effect of surface tension). A first series of experiments which led to the observation of negative-frequency waves was described in~\cite{Rousseaux:2007is}. A theoretical development of the analogy with black/white hole physics as well as rainbow physics in the context of dynamical systems theory can be found in ~\cite{Rousseaux:2009}, while~\cite{Rousseaux:2010md} contains an in-depth theoretical treatment of the dispersion relation and the associated horizon effects. Here, we limit ourselves to mentioning that there exist two possible scenarios in which the white hole horizon can be crossed:
\begin{itemize}
 \item a double-bouncing scenario in which an incoming wave bounces back at the white hole horizon, then bounces forward again at the blue horizon, after which it is sufficiently blueshifted to cross the white hole horizon;
 \item a direct dispersive penetration in which an incoming wave of sufficiently high frequency penetrates directly through the white hole horizon, in spite of the presence of a counter-current which blocks all surface waves in the gravity wave limit.
\end{itemize}
It should be noted that in the gravity wave limit, the dispersion is ``subluminal'', since the group velocity $c=\sqrt{g/k}$ decreases with $k$. The white hole horizon is then a strict one-way membrane for pure gravity waves. The inclusion of surface tension is therefore crucial for both horizon penetration scenarios just mentioned. We refer to~\cite{Rousseaux:2010md} for further details, and focus on the shallow-water case of the circular jump in the remainder of this text.

\section{The circular hydraulic jump}
When a vertical fluid jet impacts on a horizontal plate with a sufficient flow rate, it will form a thin layer near the impact zone, which expands radially and at a certain distance forms a sudden circular hydraulic jump. 

\subsection{Mach cone experiment}
\begin{figure}
\includegraphics[width=.42\textwidth, height=22ex]{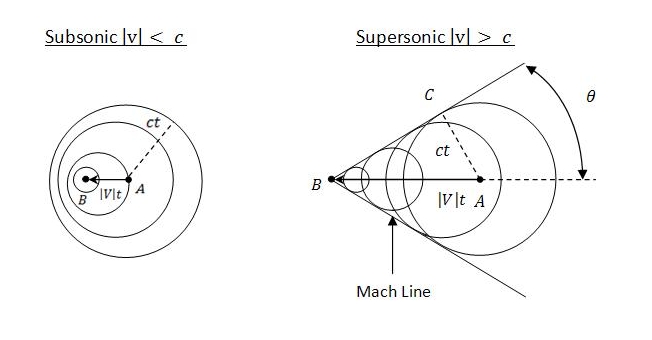}
\hspace*{-2ex}\includegraphics[width=.24\textwidth,height=18ex]{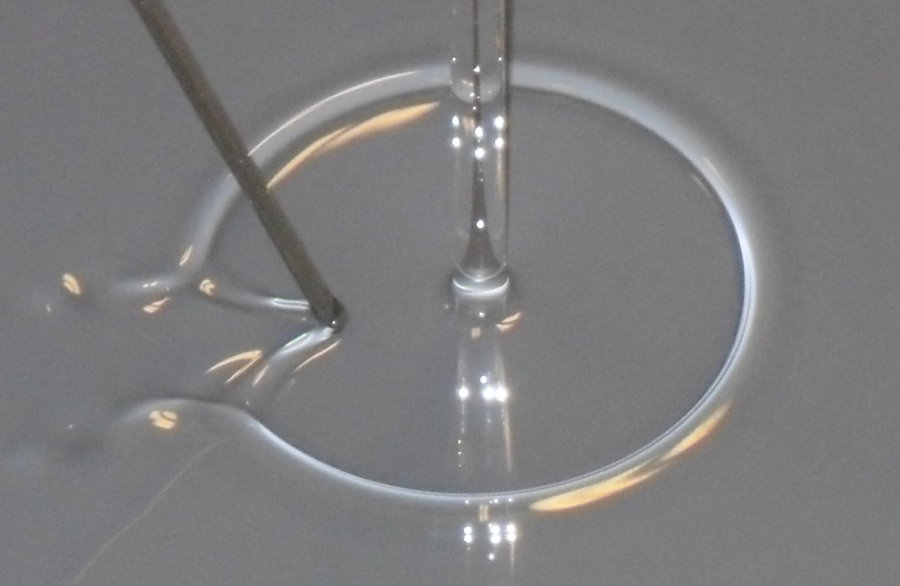}
\includegraphics[width=.165\textwidth,height=13.5ex]{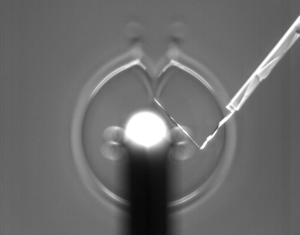}
\includegraphics[width=.165\textwidth,height=13.5ex]{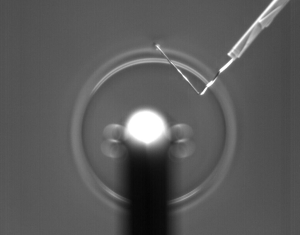}

\caption{\label{Fig:Mach-cone-theory} 
From left to right: Wavefronts emitted in subsonic/subcritical regime. Formation of Mach cone in supersonic/supercritical regime. Mach cone in circular jump (our experiments). Mach cone in circular jump, photo taken with high-speed camera. Disappearance of Mach cone just outside the jump.
}
\end{figure}

We have focused on the behaviour of surface waves propagating inward against the fluid flow of the jump itself, and in particular on the question of whether and where they are blocked. Since the propagation of surface waves on the circular jump can be described in terms of an effective Painlev\'e-Gullstrand metric, the location where the surface waves are blocked forms the hydrodynamical analogue of a white hole horizon. The above question could be answered by comparing the radial fluid velocity $v_s^r$ at the surface and the propagation velocity $c$ of the surface waves. However, given that there exist no satisfactory simultaneous measurements of both quantities, we have opted for a simpler alternative, which allows us to directly determine the ratio between them. Our method is based on the Mach cone well known in the case of sound waves, see Fig.~\ref{Fig:Mach-cone-theory}. There, an object propagating at a speed $V$ above the speed of sound $c_s$ leaves behind an observable cone, the Mach cone. This is formed by the envelope of the subsequent wavefronts emitted by the object, which partially escape from each other. The half-angle $\theta$ of the cone obeys $\sin \theta=ct/|V|t=c/|V|=1/M$, with $M$ the Mach number, see Fig.~\ref{Fig:Mach-cone-theory}. For an object propagating at a speed $V<c_s$, the subsequent wavefronts remain inside the previous ones and no Mach cone is formed. 

Exactly the same argument can be applied to the case of an object standing still at the surface of a fluid flow. By comparing the fluid flow velocity $v_s^r$ to the propagation speed of surface waves $c$, the following cases can occur:
\begin{itemize}
 \item Supercritical region: $v_s^r>c ~~\Rightarrow~~ \theta \in \left[0,\frac{\pi}{2}\right]$
 \item Subcritical region: $v_s^r<c ~~\Rightarrow~~ \theta$ complex; Mach cone disappears
 \item Horizon: $v_s^r=c ~~\Rightarrow~~ \theta=\frac{\pi}{2}$
\end{itemize}

\begin{figure}
\begin{center}
\includegraphics[width=.42\textwidth,height=30ex]{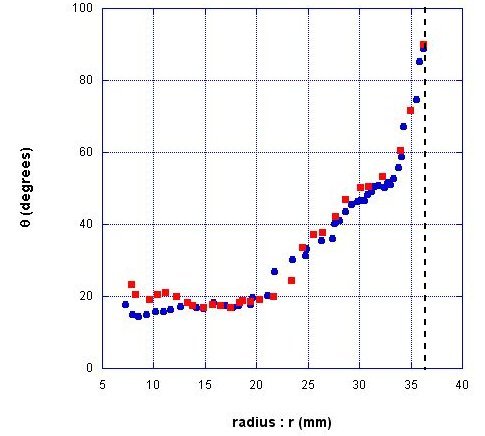}
\includegraphics[width=.4\textwidth,height=30ex]{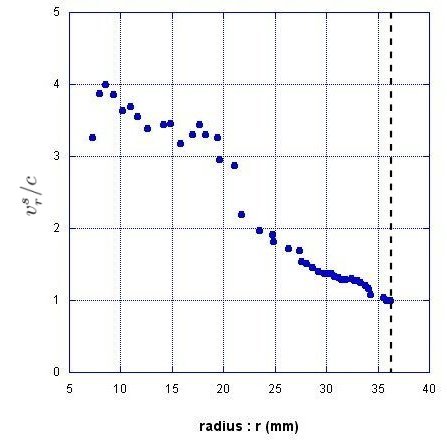}
\end{center}
\caption{\label{Fig:results-mach-angle} 
Left: Mach angle $\theta$ as a function of the distance $r$ from the centre of the jump for two values of the external fluid height $H$ (blue circles: $H=0$mm, red squares: $H=3$mm, respectively). Right: $v_ r^s/c$ as a function of $r$. The dashed vertical line represents the jump radius $R_j$. Experimental parameters: see~\cite{Jannes:2010sa}.}
\end{figure}

Our experiment to demonstrate the presence of a hydrodynamic horizon is described in~\cite{Jannes:2010sa}. Essentially, we have pumped silicon oil through a steel nozzle onto a horizontal PVC plate. A needle was placed such as to penetrate the flow surface at varying  distances from the centre of the circular jump, where the oil jet impacts on the PVC plate. For each position of the needle, we haved photographed the setup with a high-speed camera and measured the corresponding Mach angle $\theta$. The resulting angles and the derived ratio $v_ r^s/c$ are shown in Fig.~\ref{Fig:results-mach-angle}.

These results provide a clear proof that the circular hydraulic jump constitutes a two-dimensional hydrodynamic white hole: surface waves travelling at a velocity $c$ from outside the jump are trapped outside the jump in precisely the same sense as light is trapped inside a gravitational black hole. The corresponding white-hole horizon is situated precisely at the radius of the jump itself. The following features of this hydrodynamic white hole are particularly striking, especially in comparison with other current or planned experiments, e.g. in optics or Bose-Einstein condensates:
\begin{itemize}
 \item The white hole here is created ``spontaneously''. One only needs to arrange for a sufficient fluid flow rate, but no extraordinary engineering is required. In fact, it suffices in principle to open the tap in a kitchen sink to observe such a white hole. All other ingredients in the experimental setup (the choice of silicon oil, the precision of the pump etc) serve merely to make the experiment cleaner and free of perturbations and other undesired effects, but do not affect the essential point which is the creation of the hydrodynamic white hole.
 \item The white hole can be observed with the naked eye: the location of the white hole horizon is precisely where the fluid undergoes the characteristic jump.
\end{itemize}
Finally, let us go back to dispersion relation for surface waves and examine in which dispersive regime the circular jump typically lies.

\subsection{Dispersion relation}
Developing Eq.~\eqref{disp-rel} for the case $kH\ll1$ and truncating at $\mathcal{O}(k^4)$, one obtains
\begin{align*}
(\omega-Uk)^2\approx ~& gHk^2 + \left(\frac{\gamma H}{\rho}-\frac{gH^3}{3}\right) k^4+\mathcal{O}(k^6)
\\= ~& c^2k^2 + c^2\left(l_c^2-\frac{H^2}{3}\right)k^4+\mathcal{O}(k^6),
\end{align*}
where $l_c=\sqrt{\gamma/g\rho}$ is the capillary length ($l_c=1.49$mm for the silicon oil in our experiments). Contrarily to the deep-water case of the wave channel, the shallow-water regime is relativistic at low values of $k$ (the associated ``relativistic speed'' $c$ is $c=\sqrt{gH}$, i.e. the surface wave velocity in the low-$k$ or gravity limit where capillarity is negligible). Moreover, since the typical heights of the fluid inside the jump are very small (certainly smaller than $\sqrt{3}l_c$, both in our experiments and in other experiments reported in the literature), one is tempted to conclude that the circular jump should exhibit superluminal dispersion. Calculations of the the group velocity $c_g\equiv \partial \omega/\partial k$ from the complete dispersion relation confirm that the circular jump is superluminal ($c_g$ increasing with $k$) for realistic experimental parameters. 
In such a superluminal regime, sufficiently high-frequency modes can penetrate across the horizon in the classically prohibited sense. The interior of the black/white hole is then no longer causally separated from the outside. Such superluminal horizon-crossing effects are considered in several scenarios for quantum gravity phenomenology. They are particularly interesting in the sense that they put the robustness of several aspects of black hole physics (and in particular, Hawking radiation) to the test~\cite{Barcelo:2008qe}. This implies that it should be possible to test some of these issues associated to the robustness of semiclassical gravity with respect to transplanckian physics in the circular hydraulic jump.

We will report in the near future about further experiments to study the interaction between the circular jump and incoming surface waves.

\section*{References}

\providecommand{\newblock}{}

\end{document}